\documentclass[a4paper, 11pt]{article}

\usepackage[dvips]{graphicx}
\usepackage{cite}
\newcommand{\ns}{\!\!\!}

\begin{document}
\twocolumn
[
\title{\textsc{Eternal expansion of closed universe}}
\author{\textit{O. B. Karpov}\\
        \textit{{\small Moscow State Mining University,}}\\
        \textit{{\small 119991, Moscow, Russia}}}
\maketitle
\vspace{-10ex}

\renewcommand{\abstractname}{ }\abstractname
\begin{abstract}
\textsf{ The potential barrier of the closed universe expansion
has been investigated and its overcoming condition has been
obtained. The restrictions on the Friedmann integrals,
cosmological constant and medium components energy densities have
been analyzed. The phase-space has been considered and the phase
curves of eternally expanding closed universes have been plotted.
A questionable coincidence of the our Universe Friedmann integrals
has been discussed.}
\end{abstract}\\
\vspace{5.5ex} ]

\section{Introduction}

     As it is known, data on distance modulus versus redshift
reveal the presence of an acceleration of the cosmological
expansion \cite{1,2,3}. This is possible in existence of a
cosmological repulsion which is described usually by
$\Lambda$-term or equivalently by the presence of an vacuum-like
medium with a negative pressure. An open universe will expand
forever even if  $\Lambda=0$. A closed universe at $\Lambda=0$
will stop expanding in future and begin to recollapse without
fail. If $\Lambda>0$, the expansion becomes accelerated when the
doubled vacuum density exceeds the decreasing matter density. The
open universe is sure to pass this condition and its eternal
expansion has not an alternative\footnote {In the work \cite{4}
the possibilities of recollapse of a flat universe owing to decay
of cosmological constant, collision with a null singularity and
formation of space-like curvature singularity during expansion
have been investigated. In this work these possibilities are not
considered.}. Reaching of the closed universe the acceleration
state and its consequent eternal expansion requires a potential
barrier overcoming which is possible under certain conditions. In
this work a restriction on the closed universe parameters being
necessary to getting over the barrier and the universe dynamics in
this case are investigated.

\section{Condition of eternal \\ expansion}

The closed universe dynamics is described by the Einstein field
equation for curvature radius $a$ of 3-space in comoving with
cosmological expansion nonrotating (synchronous) frame:
\begin{equation}\label{eq:1}
\dot{a}^{2}=\frac{A_{m}}{a}+\frac{a^{2}}{A_{v}^{2}}-1.
\end{equation}
Here and below convention $c=1$ is used. The evolution constants
$A_{m}$ and $A_{v}$ following \cite{5, 6} are named the Friedmann
integrals. The matter integral
\begin{equation}\label{eq:2}
A_{m}=\frac{8\pi}{3}G\rho_{m}a^{3},
\end{equation}
where $\rho_{m}$ is the dust-like matter density including a dark
matter, $G$ is the gravitational constant. The vacuum integral
$A_{v}$
\begin{equation}\label{eq:3}
A_{v}^{-2}=\frac{8\pi}{3}G\rho_{v}=\frac{\Lambda}{3},
\end{equation}
where $\rho_{v}$ is the vacuum energy density, $\Lambda$ is the
cosmological constant. The equation (\ref{eq:1}) has been written
without  taking into account a radiation which contribution
dominates at an early stage of the universe evolution and is
negligible at the stage considered. Define
\[
U_{m}=-\frac{A_{m}}{a},\quad U_{v}=-\frac{a^{2}}{A_{v}^{2}}, \quad
U=U_{m}+U_{v}
\]
and write the equation (\ref{eq:1}) in form of an "energy
conservation law"
\[
\dot{a}^{2}+U(a)=-1.
\]

     Graphs of the functions $U_{m}(a)$, $U_{v}(a)$, $U(a)$ are plotted
in Fig. 1.
\begin{figure}[h]
    \centering
    \includegraphics[width=0.47\textwidth]{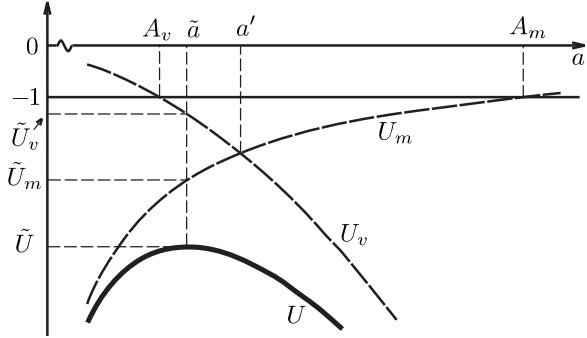}
    \caption{Potential barrier}
\end{figure}
 At the point $a=\tilde{a}$,
\begin{equation}\label{eq:4}
\tilde{a}^{3}=\frac{1}{2}A_{m}A_{v}^{2}
\end{equation}
the maximum $\tilde{U}$ of the potential barrier $U(a)$ is
disposed,
\begin{equation}\label{eq:5}
\tilde{U} = U(\tilde{a}) = -\left(\frac{\alpha
A_{m}}{A_{v}}\right)^{2/3},
\end{equation}
where $\alpha = 3\sqrt{3}/2$. At this point the balance of matter
gravitation and vacuum antigravitation is achieved, $\rho_{m} =
2\rho_{v}$. The levels $\tilde{U}$, $\tilde{U}_{m}$,
$\tilde{U}_{v}$, $0$ are equidistant:
\[
\tilde{U}-\tilde{U}_{m} = \tilde{U}_{m}-\tilde{U}_{v} =
\tilde{U}_{v} = \frac{1}{3}\tilde{U}.
\]
At the point $a=\tilde{a}$ the deceleration parameter changes the
sign:
\begin{equation}\label{eq:6}
q = -\frac{\ddot{a}a}{\dot{a}^{2}} =
\frac{\tilde{a}^{3}-a^{3}}{a^{3}-aA_{v}^{2}+2\tilde{a}^{3}}.
\end{equation}
At $a<\tilde{a}$ the matter gravitation dominates and $q>0$, at
$a>\tilde{a}$ the vacuum antigravitation dominate and $q<0$. At
the point $a=a'$,
\[
a'^{3}=2\tilde{a}^{3}
\]
$U'_{m}=U'_{v}$ and $\rho_{m}=\rho_{v}$. If the level $U=-1
> U'_{m}=U'_{v}$ then $A_{m}>A_{v}$ (as represented in Fig.~1);
$A_{m} \leq A_{v}$ otherwise. The Friedmann integral $A_{m}$ is a
maximum value of the curvature radius in the standard model
$\Lambda=0$, the integral $A_{v}$ equals an initial value $a$ in
the de Sitter universe $\rho_{m}=0$:
\begin{equation}\label{eq:7}
a(t)=A_{v}\cosh(t/A_{v}).
\end{equation}

     A closed universe will expand eternally if the level $U=-1$
is above the potential barrier summit $U=\tilde{U}$ (\ref{eq:5}),
$|\tilde{U}|\geq 1$ (the equality sign corresponds to the
asymptotic approaching $a$ to $\tilde{a}$ (\ref{eq:4})):
\begin{equation}\label{eq:8}
\frac{\alpha A_{m}}{A_{v}}\geq1.
\end{equation}
The condition of potential barrier overcoming corresponds to the
strict inequality. Thus, universe fate is determined by the
Friedmann integrals ratio. The condition (\ref{eq:8}) means the
limitation on the cosmological constant
$$
\Lambda\geq\frac{4}{3}A_{m}^{-2}. \eqno (8')
$$
The $A_{m}$ value (\ref{eq:2}) can be estimated very rough due to
a considerable uncertainty of the scale factor $a$. According to
\cite{5, 6} $A_{v}\sim A_{m}$ and even some more, and then a
fulfilment of the condition (\ref{eq:8}) turns out to be
problematic.

    Find out the condition (\ref{eq:8}) for the density parameters
$\Omega_{m}$ and $\Omega_{v}$,
\begin{eqnarray}
\Omega_{m}&\ns=\ns&\frac{\rho_{m}}{\rho_{c}} = \left(1 +
\left(\frac{a}{a'}\right)^{3} -\frac{a}{a'}
\left(\frac{A_{v}}{A_{m}}\right)^{\frac{2}{3}}\right)^{-1}, \label{eq:9}\\
\Omega_{v}&\ns=\ns&\frac{\rho_{v}}{\rho_{c}} =
\left(\frac{a}{a'}\right)^{3}\Omega_{m} = (HA_{v})^{-2},
\label{eq:10}
\end{eqnarray}
where $\rho_{c}$ is the critical density, $\rho_{c}=3H^{2}/8\pi
G$, $H=\dot{a}/a$ is the Hubble constant. The density parameter
$\Omega_{m}$ achieves a maximum at $a=a_{1}$
\[
a_{1}=\frac{A_{v}}{\sqrt{3}};
\]
$\Omega_{m}=1$ at $a=0$ and $a=A_{v}$. The parameter $\Omega_{v}$
achieves a maximum (and the Hubble constant $H$ - a minimum) at
$a=a_{2}$
\[
a_{2}=\frac{3}{2}A_{m};
\]
$\Omega_{v}=1$ at $a^{-1}=0$ and $a=A_{m}$. The deceleration
parameter (\ref{eq:6})
\begin{equation}\label{eq:11}
q=\frac{1}{2}\Omega_{m}-\Omega_{v};
\end{equation}
$q=0.5$ at $a=0$ and $a=a_{1}$, $q=-1$ at $a^{-1}=0$ and
$a=a_{2}$. The curvature radii $a_{1}$, $a_{2}$ and $\tilde{a}$
satisfy the ratios
\[
\frac{a_{2}}{a_{1}} = \left(\frac{a_{2}}{\tilde{a}}\right)^{3/2} =
                      \left(\frac{\tilde{a}}{a_{1}}\right)^{3} =
                      \frac{\alpha A_{m}}{A_{v}}.
\]
The Einstein field equation (\ref{eq:1}) connects the density
parameters and the Friedmann integrals by the correlation
\begin{equation}\label{eq:12}
\left(\Omega_{m}+\Omega_{v}-1\right)^{3} =
\left(\frac{A_{v}}{A_{m}}\right)^{2}\Omega_{m}^{2}\Omega_{v}.
\end{equation}
The condition of the eternal expansion of the closed universe
(\ref{eq:8}) means
\begin{equation}\label{eq:13}
\left(\Omega_{m}+\Omega_{v}-1\right)^{3} \leq
\alpha^{2}\Omega_{m}^{2}\Omega_{v}.
\end{equation}
If in accordance with the last data \cite{7} we assume
$\Omega\simeq 0.03$, where $\Omega=\Omega_{m}+\Omega_{v}$, then
the equation (\ref{eq:12}) yields $A_{v} \sim 10^{-2}A_{m}$ and
the condition (\ref{eq:8}), (\ref{eq:13}) is fulfilled easily.

    As it is clear from (\ref{eq:12}), the ratio $A_{v}/A_{m}$
is connected with the degree of the universe spatial flatness, and
$A_{v}/A_{m}\rightarrow 0$ when $\Omega\rightarrow 1$. Therefore
the problem of the cosmic coincidence of the Friedmann integrals
$A_{m}$ and $A_{v}$ discussed in the works \cite{5, 6} seems to be
not actual. Do not coincide the expansion factor $a$ today with
constant $A_{v}$ either:
\begin{equation}\label{eq:14}
\left(\frac{a}{A_{v}}\right)^{3} = \frac{\Omega_{v}}{\Omega_{m}}
\frac{A_{m}}{A_{v}}
\end{equation}
and $a/A_{v}\rightarrow\infty$ when $\Omega\rightarrow 1$. There
exists the problem of coincidence of $\Omega_{m}$ and $\Omega_{v}$
that should be explained by anthropic arguments \cite{3, 8}.

\section{Phase curves}

Curves in Fig. 2 represents the dynamics of universes according to
(\ref{eq:12}).
\begin{figure}[b]
    \centering
    \includegraphics[width=0.47\textwidth]{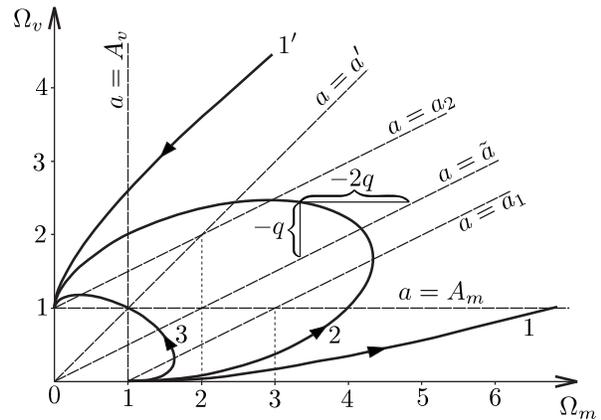}
    \caption{Phase curves}
\end{figure}
 The point $(1, 0)$ corresponds to a Big Bang, the
point $(0, 1)$ signifies a de Sitter universe. The evolution of an
universe with the fixed Friedmann integrals ratio $A_{m}/A_{v}$ is
described by a the curve in $(\Omega_{m}, \Omega_{v})$ plane.  One
and only one curve passes through each of the point $(\Omega_{m},
\Omega_{v})$, excepting $(0,1)$ and $(1,0)$. It means an
observational determination of the values $\Omega_{m}$,
$\Omega_{v}$ fixes the universe evolution in the model
circumscribed (\ref{eq:1}), (\ref{eq:12}). The phase-space area
$(\Omega_{m}, \Omega_{v})$ satisfying the condition (\ref{eq:8}),
(\ref{eq:13}) is disposed between the curves 1 and $1'$ which
correspond to the equality $A_{v}=\alpha A_{m}$.  All the curves
in this area corresponding the universes expansion begin at the
point $(1, 0)$; the evolutions are completed by the de Sitter
universe $(0, 1)$. On the curves 2 and 3 the correlations $A_{v}=2
A_{m}$ and $A_{v}=A_{m}$ are fulfilled respectively. Below the
curve 1 universes expand from the Big Bang $(1,0)$ till a maximum
value $a$ defined left point of intersection of the straight line
$U=-1$ and the curve $U(a)$ when $\Omega_{m}$ and $\Omega_{v}$ go
to infinity, afterwards the universes begin to recontract and
recollapse $(1,0)$. Above the curve $1'$ expansion goes on the
other (large $a$) side of the potential barrier without the Big
Bang from a minimum value $a$ defined the right intersection point
of the lines $U=-1$ and $U(a)$ to the de Sitter metric $(0,1)$. A
passage from $(1, 0)$ to $(0, 1)$ on a straight line corresponds a
flat universe $\Omega_{m}+\Omega_{v}=1$. The dependence $a(t)$
(\ref{eq:7}) at $t\rightarrow\infty$ becomes exponential
\[
a(t)=A_{v}\exp(t/A_{v}),
\]
which describes the final fate of any expanding universe with the
$(\Omega_{m}, \Omega_{v})$ values above the curve 1.

    Plots of the parameters $\Omega_{m}$ (\ref{eq:9}),
$\Omega_{v}$ (\ref{eq:10}), $q$ (\ref{eq:11}) versus $a/\tilde{a}$
for the universe $A_{v}=2 A_{m}$ (the curve 2 in Fig. 2) and for
the almost flat universe $A_{v} = 10^{-2}A_{m}$ are given in Fig.
3(a) and 3(b) respectively.
\begin{figure}[h]
    \centering
    \includegraphics[width=0.45\textwidth]{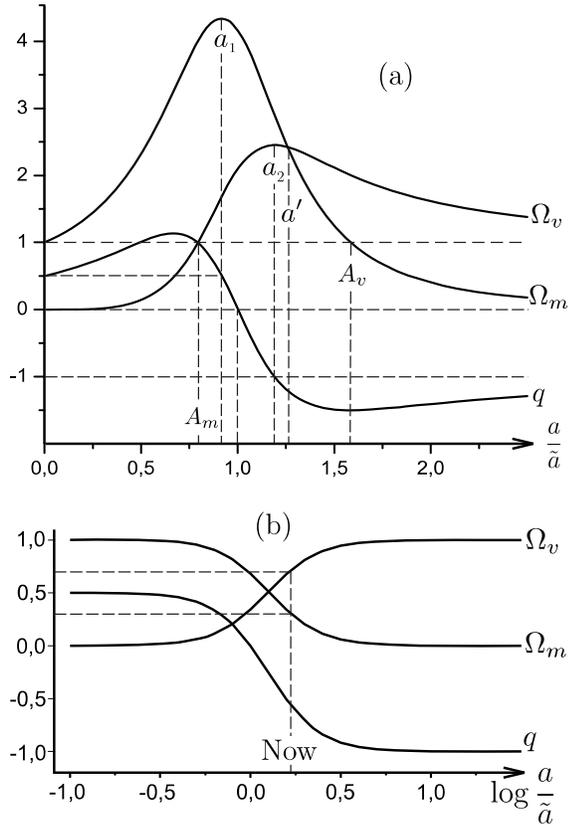}
    \caption{Universes dynamics.
   (a) $A_{v}=2A_{m}$. (b) $A_{v}=10^{-2}A_{m}$}.
\end{figure}

    It should be mentioned that at an early stage of an universe
evolution a radiation density $\rho_{r}$ dominates, then
phase-space is three-dimensional $(\Omega_{m}, \Omega_{v},
\Omega_{r})$. The phase curves begin at the point $(0,0,1)$; they
are directed to the $\Omega_{r}$ axis for a closed universe, and
when the matter grows predominant over the radiation, the curves
pass on $(\Omega_{m}, \Omega_{v})$ plane as it is shown in Fig. 2.
The vacuum energy density $\rho_{v}$ is to prevail at a still more
early stage preceding radi\-ation-dominated, and then the universe
develops from de Sitter metric to de Sitter metric along a closed
phase curve. This curve simplified has been plotted here for a
flat universe $\Omega_{m}+\Omega_{v}+\Omega_{r}=1$.
\begin{figure}[h]
    \centering
    \includegraphics[width=0.18\textwidth]{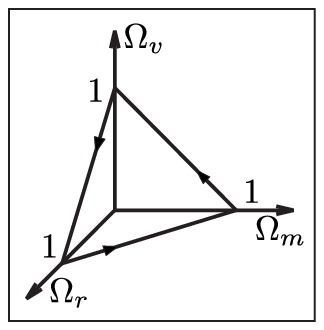}
\end{figure}

\section{Summary}

    The measured cosmological expansion acceleration means that our
Universe will expand eternally. The condition of the closed
universe eternal expansion restricts the Friedmann integrals ratio
by the correlation (\ref{eq:8}). Written down for the density
parameters $\Omega_{m}$, $\Omega_{v}$ this correlation
(\ref{eq:13}) separate the area in the phase-space (Fig. 2). The
measured parameters $\Omega_{m}$, $\Omega_{v}$ of our Universe are
found in this area. The Friedmann integrals of our Universe do not
coincide. The phase curves of the eternally expanding closed
universes have been plotted. The dependence of the cosmological
parameters $\Omega_{m}$, $\Omega_{v}$, $q$, $H$ on the universe
curvature radius $a$ has been investigated.

\end{document}